\documentclass[showpacs,prb,amsmath,twocolumn,floatfix]{revtex4}

\usepackage{subfigure}
\usepackage{graphicx}
\usepackage{amssymb}

\begin{document}

\title{Stable fractional flux vortices in mesoscopic superconductors}

\author{L.F. Chibotaru and V.H. Dao}
\affiliation{Division of Quantum and Physical Chemistry and INPAC - Institute for Nanoscale Physics and Chemistry, Katholieke Universiteit Leuven,
Celestijnenlaan 200F, B-3001 Leuven, Belgium}


\begin{abstract}

Conventional superconductors have vortices carrying integer multiples of magnetic flux quantum while unconventional ones, with $p$- or $d$-wave order parameter, allow half-integer fluxes.  
Here we show that mesoscopic size effects stabilize fractional flux vortices in the {\em thermodynamical ground state} of $s$-wave two-gap superconductors. The value of these fluxes can be an {\em arbitrary fraction of flux quantum} and can be measured directly from distributions of magnetic fields on the samples.

\end{abstract}


\maketitle       

The quantization of magnetic fluxes associated to vortices is one of the most direct demonstrations of the existence of macroscopic quantum coherence in superconductors.
The phase of superconducting order parameter (OP) can change only by multiples of $2 \pi$ along any closed contour in ordinary superconductors, which implies that the magnetic flux threading the vortex is quantized in units of $\Phi_0 =hc/2|e|$. In contrast, the spin degree of freedom in triplet OP topologically allows half-quantum vortices \cite{volovik,leggett}. However the energy of these vortices diverges in bulk samples due to unscreened spin currents \cite{sigrist}. Theoretical arguments were given for their stabilization by formation of pairs or by mesoscopic size effects, where the divergent part of the energy is removed \cite{chung}. However the question whether such phases could be realized as true thermodynamical ground state remains open. Half-quantum vortices have been also predicted and actually observed in heterostructures by combining $d$-wave superconductors with superconductors of other symmetries, which resulted in $\pi$-Josephson junctions \cite{geshkenbein,harlingen,hilgenkamp}.

Another type of fractional flux vortex was suggested in the context of multicomponent/multiband superconductors \cite{tanaka,babaev2}.
Multicomponent models gained recently much attention in connection with 
the discovery of two-gap superconductivity in MgB$_2$ \cite{nagamatsu} and iron pnictides \cite{kamihara}, where they are expected to exhibit a plethora of new properties \cite{babaev,moshchalkov,gurevich1,blumberg,babaev1,chibotaru}, and also appear in the description of the superfluidity of liquid metallic hydrogen or deuterium \cite{babaev1} and  Bose-Einstein condensates of dilute atomic gases \cite{kasamatsu}. 
Fractional vortices in these systems are quite different from those arising in unconventional superconductors since they are neither bound to interfaces, like the $\pi$-junction vortices, nor are they related to a special symmetry of OP. However, as in the case of unconventional superconductors, such vortex phases are energetically forbidden in the bulk \cite{babaev2}. This restriction is in full accord with the fact that no fractional flux vortices have been observed to date in all investigated multicomponent superconductors. Contrary to that, we prove here that fractional flux vortices exist in mesoscopic two-component superconductors as thermodynamically stable phases in a broad range of temperatures, applied magnetic fields and superconducting parameters. Moreover, our calculations show that fluxes associated to these vortices can take arbitrary fractions of $\Phi_0$ and should be observable with existing measuring techniques of magnetic field distributions.

For arbitrary (non-periodic) vortex patterns, the flux associated to each vortex threads the domain confined by a contour $\Gamma$ which obeys the relation
\begin{equation}
\oint_{\Gamma} \frac{{\mathbf j}({\mathbf l}) d{\mathbf l}}{|\Psi ({\mathbf l}) |^2} =0 , 
\label{j}
\end{equation}
where $\Psi ({\mathbf r})$ is the Ginsburg-Landau (GL) OP and ${\mathbf j}({\mathbf r})$ is the current density. Such definition of $\Gamma$ automatically implies the quantization of flux associated to a vortex in conventional superconductors \cite{comment}.
In mesoscopic superconductors, the quantization of the flux associated to a vortex is only defined when the contour $\Gamma$ lies entirely within the sample. When this condition is not fulfilled \cite{unpublished}, the fractionalization of this flux arises in conventional superconductors \cite{bardeen,geim} as a simple consequence of the fact that it has not entered yet entirely in the sample. 
Note that even if the fluxes of individual vortices are quantized, the total flux threading the mesoscopic sample will be not. This is because a non-quantized flux passes through the domain between the contour(s) $\Gamma$ and the boundary of the sample, where the current flow is diamagnetic.  To be distinguished from these situations, here we consider the case of fractionalization of \textit{the flux associated to a vortex when it has entered completely} in the superconductor, i.e. the contour $\Gamma$ lies entirely within the sample. 
  
The current density for the two-component superconductor is given by the GL expression  
${\mathbf j}= -2e\hbar [\text{Im} (\Psi_1^* \nabla \Psi_1 )/m_1 +\text{Im} (\Psi_2^* \nabla \Psi_2 )/m_2 ] -4(e^2 /c)(|\Psi_1 |^2 /m_1+ |\Psi_2 |^2 /m_2 ) {\mathbf A} $, where $\Psi_1$ and $\Psi_2$ are the  OPs in the two condensates, $-2e$ is the charge of the Cooper pair and ${\mathbf A}$ is the vector potential corresponding to the magnetic induction ${\mathbf b}({\mathbf r})$. Choosing a contour $\Gamma$ satisfying the relation
\begin{equation}  
\oint_{\Gamma} \frac{{\mathbf j}({\mathbf l}) d{\mathbf l}}{\frac{|\Psi_1 ({\mathbf l}) |^2}{m_1} + \frac{|\Psi_2 ({\mathbf l}) |^2}{m_2} }=0 ,
\label{j1}
\end{equation}
we obtain for the magnetic flux threading the region confined by $\Gamma$:
\begin{equation}  
\Phi_\Gamma = -\frac{1}{2\pi}\oint_{\Gamma} \frac{\frac{\text{Im} [\Psi_1 ({\mathbf l})^* \nabla \Psi_1 ({\mathbf l})]}{m_1} +\frac{\text{Im} [\Psi_2^* ({\mathbf l})\nabla \Psi_2 ({\mathbf l})]}{m_2} } {\frac{|\Psi_1 ({\mathbf l}) |^2}{m_1} + \frac{|\Psi_2 ({\mathbf l}) |^2}{m_2} }d{\mathbf l} \;\Phi_0 .
\label{flux}
\end{equation}
For $\Psi_2 ({\mathbf r}) \propto \Psi_1 ({\mathbf r})$ Eq. (\ref{flux}) reduces to the familiar flux quantization, while in the case of a composite vortex with winding numbers $L_1$ and $L_2$ in the two condensates, respectively, we obtain in the London limit \cite{babaev2}:
$\Phi_\Gamma = (L_1 |\Psi_1 |^2 /m_1 + L_2 |\Psi_2 |^2 /m_2 )/(|\Psi_1 |^2 /m_1 + |\Psi_2 |^2 /m_2 ) \Phi_0 $.
The last expression shows that the flux can become fractional if $L_1 \neq L_2$, which, however, is never the case in bulk superconductors as mentioned above.

Consider a long superconducting cylinder of radius $R\sim \xi (0)$ in external homogeneous field $H$ applied along its axis, described by a two-band GL free energy functional \cite{dao1}: 
\begin{equation} 
F=\sum_{n=1}^{2} F_n + \int\big[ 
\frac{1}{8\pi}({\mathbf b}-{\mathbf H})^2 -\gamma (\Psi_1^* \Psi_2 + \Psi_1 \Psi_2^* ) \big] dV, 
\label{F}
\end{equation}
%
%
%
where the single-band functionals $F_n=\int \big( \alpha_n |\Psi_n |^2 + \beta_n |\Psi_n |^4/2 + |\Pi \Psi_n |^2/2m_n  \big) dV$ with $\Pi =-i\hbar\nabla +\frac{2e}{c}{\mathbf A}$ are supplemented with  the magnetic energy and the Josephson-type coupling of the two condensates. For long cylinders the problem reduces to two dimensions and the solutions are presented in the form: $\Psi_1 = \sqrt{|\alpha_1 |/\beta_1 } \sum_L u_L (r/R) \text{e}^{iL\theta}$, $\Psi_2 = \sqrt{|\alpha_1 |/\beta_1 } \sum_L v_L (r/R) \text{e}^{iL\theta}$. The stable state is found by full numerical minimization \cite{numerical} of the GL functional with respect to $u_L$, $v_L$ and ${\mathbf A}$ in each point of space under the boundary conditions $\Pi \Psi_1 |_\text{n} =0$, $\Pi \Psi_2 |_\text{n} =0$, and ${\mathbf b}|_\text{t} =H$ at $r=R$.

\begin{figure}
\begin{center}
 \scalebox{0.95}{%
 \includegraphics*{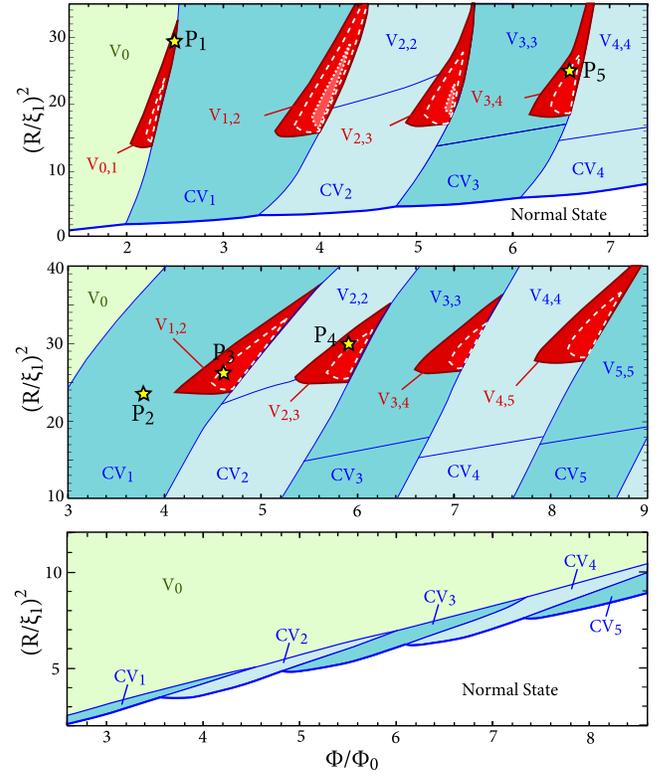}}
\end{center}
\caption{(color online) Phase diagram of the two-component superconducting cylinder for $\kappa = 10$ (top), 3 (middle) and 0.5 (bottom) ($x_{20}= 12$, 20 and 0.5, respectively) with $x_\gamma$ =0.01. It is divided into domains of superconducting states with no vortex (V$_0$), with a central giant vortex of winding number $L$ in each condensate (CV$_L$), and with $L_n$ separated vortices in the $n$-th condensate (V$_{L_1,L_2}$). Dashed and dotted lines delineate the domains where fractional flux vortices exist as stable phases for $x_\gamma$ =0.05 and 0.1, respectively. The stars on the plots denote points at which current distributions are shown in Fig. 3.}
\end{figure}

We will assume that the first band is dominant, i.e. $\alpha_2$ becomes negative at a temperature lower than $\alpha_1$. The coherence length is related to temperature via $\xi_1^2= -\hbar^2/2m_1 \alpha_1 $ and $\alpha_2/|\alpha_1|= x_{20} (\xi_1/R)^2 - \beta_2/\beta_1$ where $x_{20}$ is a constant that quantifies the difference of intrinsic critical temperatures between the two bands. It is then convenient to parametrize the phase diagram with $(R/\xi_1)^2$ and $\Phi =\pi R^2 |H|$. Fig. 1 shows the calculated phase diagram for $m_1 =m_2$ and $\beta_1 = \beta_2$  at different values of GL parameter $\kappa$ ($\equiv\ \kappa_1 =\kappa_2 =\Phi_0 m_1 \beta_1^{1/2}/(2\pi)^{3/2}\hbar^2$) and interband coupling $x_\gamma =\gamma /(\hbar^2 /2m_1 R^2)$. At high temperature (close to the $T_c(H)$ boundary line) the OPs are characterized by a giant vortex in the center of the cylinder ($L_1 =L_2$) much as in the case of conventional superconducting cylinders \cite{saintjames,moshchalkov1}. We can see that a cylinder of mesoscopic radius is an effectively type-II superconductor (vortices penetrate it) even for $\kappa <1/\sqrt{2}$ (Fig. 1, bottom), like in superconducting thin films \cite{pearl}. However for values $\kappa \lesssim 0.35\div 0.05$ the whole superconducting domain on the phase diagram is occupied by the Meissner phase ($V_0$). The shaded closed regions ($V_{L_,L+1}$) in Fig. 1 correspond to domains with fractional flux vortices as stable thermodynamical phases. Fig. 2 shows indeed that the phases with fractional flux vortices are the lowest in energy in the corresponding domains. As seen in Fig. 1, these domains shrink with increasing $\gamma$ and decreasing $\kappa$. 

\begin{figure}
 \scalebox{0.8}{%
 \includegraphics*{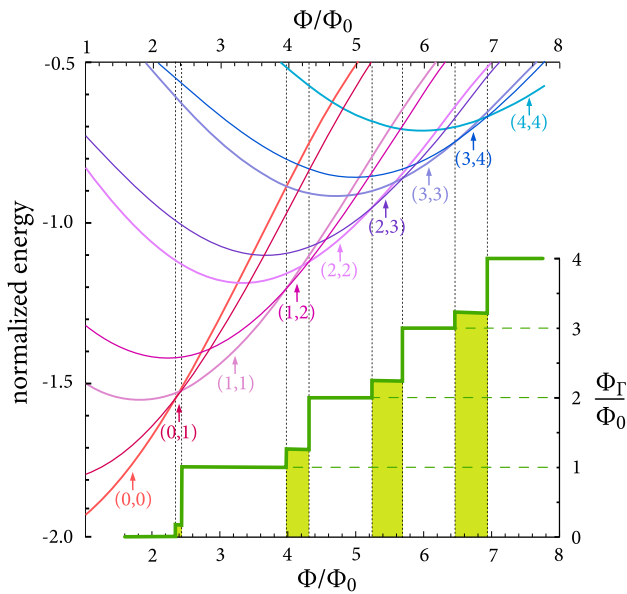}}
\caption{ (color online) Top: the free energy of the two-component superconducting cylinder for different symmetries of the OP for $\kappa =10$ and $R^2 /\xi_1^2 =25$. Vertical dot lines delineate the domains with fractional flux vortices in the stable thermodynamic phase.
Bottom: The flux associated to vortices ($\Phi_\Gamma $) in the stable thermodynamic phase.}
\end{figure}

In order to calculate the flux associated with vortices $\Phi_\Gamma $, we find first the contour $\Gamma$ using the calculated current density distributions (Fig. 3 c-f) and Eq. (\ref{j1}). To this end we draw a closed line through the points where ${\mathbf j}({\mathbf r})$ has zero tangential component. Then $\Phi_\Gamma $ is found either by applying Eq. (\ref{flux}) or by integrating the magnetic induction threading the contour.
The resulting $\Phi_\Gamma $ is shown in the lower panel of Fig. 2 as function of applied flux. We can see from the figure that in the domains where the $\Phi_\Gamma $ is fractional, it almost does not varies with the field, looking like a step of intermediate (fractional) height. The height of these steps evolves monotonically from zero at some higher temperature till some fractional value at T=0. The overall picture is shown in Fig. 4. It shows, in particular, the existence of finite domains on the phase diagram where the fractional flux varies monotonically in a broad range of values. Thus continuous changes of fractional flux could be detected via the measurements of the field distributions at different temperatures. 

\begin{figure}
 \scalebox{0.95}{%
 \includegraphics*{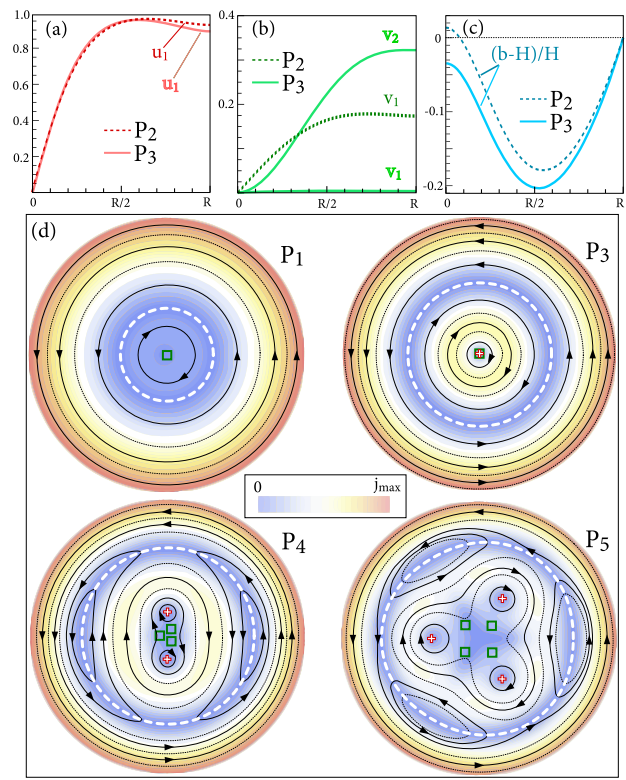}}
\caption{ (color online) OP components $u_L$ (a) and $v_L$ (b) for $R^2 /\xi_1^2 =24$ (dashed line) and 27 (solid line), corresponding to quantized and fractional vortex phases. (c) Radial variation of the magnetic induction in the corresponding phases. (d) Current density distributions and streamlines for values of $R^2 /\xi_1^2$ indicated by stars on Fig. 1. The position of vortices in the two condensates are denoted by crosses for the first band and by squares for the second band. The contour $\Gamma$ is shown by the white dashed  line.} 
\end{figure}

To understand the reason for the fractionalization of $\Phi_\Gamma $ we investigate the OPs $\Psi_1$ and $\Psi_2$ in different regions of the phase diagram (Figs. 3a,b). At a high temperature, (dashed line) the two OPs correspond to the giant vortex with the same vorticity ($L=1$) and moreover are proportional to each other. Then according to Eq. (\ref{flux}) we obtain $\Phi_\Gamma /\Phi_0 =1$, i.e. the vortex flux is quantized. For temperature low enough the giant vortex in the second band mostly corresponds to $L=2$, while the active band still contains the same giant vortex with $L=1$. Given this difference in the two OPs, Eq. (\ref{flux}) will result in a fractional value of $\Phi_\Gamma $. The current distribution for this state is shown in Fig. 3d.

The Josephson-type interaction between two condensates favors maximal overlap of $\Psi_1 ({\mathbf r})$ and $\Psi_2 ({\mathbf r})$. The two condensates also interact indirectly through the magnetic induction. When expressed with normalized quantities, it can be shown to the first order in $\kappa^{-2}$ that
$\tilde{F}= \sum_n \tilde{F}_n[\tilde{\Psi}_n,\tilde{\bf{A}}_0] - \tilde{\gamma} \int(\tilde{\Psi}_1^* \tilde{\Psi}_2 + \tilde{\Psi}_1 \tilde{\Psi}_2^*) - \kappa^{-2} \int\!\! \int \tilde{j} ({\bf r}) \tilde{j}({\bf r}')/|{\bf r}-{\bf r}'|$,
%
%
where $\tilde{F}_n[\tilde{\Psi}_n,\tilde{\bf{A}}_0]$ is the functional of the $n$-th condensate in the unperturbed applied field with the vector potential $\tilde{\bf A}_0$. The last term in the normalized functional favors the maximal overlap of the current distributions in the two condensates, i.e. the maximal similarity of the two OPs. This explains why the domains of the existence of fractional flux vortices shrink with the increase of $\gamma$ ($x_\gamma$) and the decrease of $\kappa$ (Fig. 1). 

For MgB$_2$, the most investigated two-band superconductor, the Josephson coupling is moderately weak: the smallest estimate is $\gamma \gtrsim 0.3 |\alpha_1|$\cite{chibotaru,walte} which yields $x_{\gamma} \gtrsim 0.3 (R/\xi_1)^2$. Our numerical minimizations have not found any flux fractionalization for $x_{\gamma} \gtrsim 0.2$ or $(R/\xi_1)^2 \lesssim 3$. These limitations mean that $\gamma$ in MgB$_2$ is at least five times too large. However there is no reason why $\gamma$ cannot be lower in other multiband superconductors. Among the promising candidates we can cite the heavy fermion compound UNi$_2$Al$_3$ \cite{jourdan}, the quaternary borocarbide YNi$_2$B$_2$C \cite{mukhopadhyay} or the ternary-iron silicide Lu$_2$Fe$_3$Si$_5$ \cite{nakajima} where stronger deviations from the predictions for a single gap suggest that their interband couplings may be weaker than in MgB$_2$.  

The flux fractionalization can be understood as resulting from the combination of two weakly interacting condensates which, when uncoupled, individually prefer states of different vorticities for the same applied magnetic field. So this mechanism is not restricted to the particular sample symmetry studied here and should be present in other geometries too. For example the unlocking of the condensate vorticities is not specific to long cylinders since it exists as well in the opposite limit of thin disks. We have furthermore checked that it can also occur in cylinders with different central holes \cite{unpublished}.      
   
We notice from Fig. 3 that the contour $\Gamma$ is circular for central vortex phases and is still very close to a circular shape for OPs characterized by low vorticity, despite the fact that the vortices are deconfined (vortices in the two condensates are spatially separated). The distribution of the magnetic induction for these states (Fig. 3c) shows a characteristic depletion at a radius which coincides exactly with the position of the contour $\Gamma$. Therefore for experimental determination of $\Phi_\Gamma $ in these states it would be sufficient to integrate the measured distribution of ${\mathbf b}({\mathbf r})$ in the region confined by this depletion. The amplitude of the variation of magnetic induction is ca 5 \% for $\kappa =10$ and ca 20 \% for $\kappa =3$, which is larger than the accuracy of the current measuring techniques. For phases of higher vorticity or samples of non-cylindrical geometry the contour $\Gamma $ is not circular anymore. In such cases the distribution of currents should be obtained first from the measured distribution of magnetic induction on the top surface using the relation ${\mathbf j}({\mathbf r})=(c/4\pi )\nabla\times{\mathbf b}({\mathbf r})$. Then the contour $\Gamma$ can be found as the line on which the current has no tangential component.    

In conclusion, we have proven by exact calculations within GL theory that fractional vortices, carrying arbitrary fraction of flux quantum (Fig. 4), exist in mesoscopic two-component superconductors as stable thermodynamic phases in a broad range of superconducting parameters, temperature and applied magnetic field. The origin of stabilization of these vortex phases is the different effect of the confinement on the two condensates exerted by the boundary of the sample, i.e. is a pure mesoscopic effect. This effect is counteracted by the Josephson coupling between the two condensates and the magnetic field induced by the screening currents, both favoring similar distributions of the OPs. The obtained fractional flux vortices should be observable with available experimental techniques for the investigation of magnetic field distribution.

\begin{figure}
 \scalebox{0.95}{%
 \includegraphics*{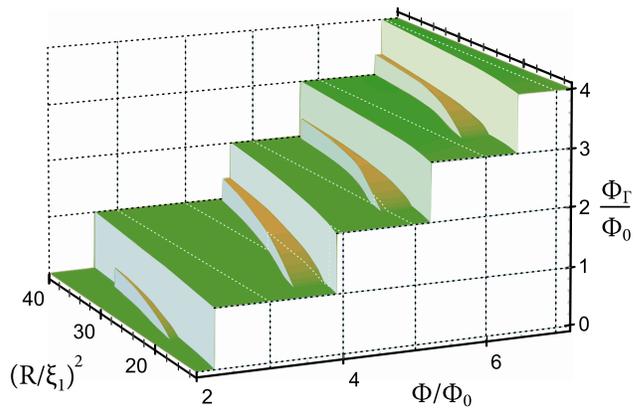}}
\caption{ (color online) The flux associated to vortices ($\Phi_\Gamma $) corresponding to the phase diagram in the upper plot of Fig.1. Domains corresponding to fractional vortex fluxes are clearly seen as nonhorizontal intermediate (noninteger) steps.} 
\end{figure}

This work has been supported by the Belgian Science Foundation and Flemish Government under 
Concerted Action Scheme. VHD acknowledges the financial support by the grant EF/05/005 (INPAC) from the 
University of Leuven.

\end{document}